\documentclass[12pt]{article}

\usepackage[dvipdfmx]{graphicx}
\usepackage{amssymb,natbib,mathtools,amsmath,setspace}
\usepackage{dcolumn,lscape,here,comment,mathrsfs,lineno}

\usepackage[dvips]{color}
\newcommand{\bm}[1]{\mbox{\boldmath $#1$}}
\newcommand{\bmt}[1]{\mbox{\tiny\boldmath $#1$}}
\newcommand{\logit}{\mathop{\rm logit}}
\newcommand{\expit}{\mathop{\rm expit}}

\newcommand{\argmin}{\mathop{\rm argmin}}

\newcommand{\sign}{\mathop{\rm sign}}

\def\department#1{\def\@department{#1}}

\title{Robust minimum divergence estimation\\ in a spatial Poisson point process}
\author{Yusuke Saigusa$^1$, Shinto Eguchi$^2$ and Osamu Komori$^3$}

\date{}

\begin{document}
\bibliographystyle{jae.bst}
\maketitle

\begin{tabular}{p{36em}}
$^1$Department of Biostatistics, School of medicine, Yokohama City University, 3-9 Fukuura, Kanazawa, Yokohama, Kanagawa 236-0004, Japan\\
$^2$The Institute of Statistical Mathematics, 10-3 Midori-cho, Tachikawa, Tokyo 190-8562, Japan\\
$^3$Department of Computer and Information Science, Seikei University, 3-3-1 Kichijojikitamachi, Musashino, Tokyo 180-8633, Japan
\end{tabular} \vspace{3pt} \\

\

\noindent
{\large \bf Abstract}

Species distribution modeling (SDM) plays a crucial role in investigating habitat suitability and addressing various ecological issues.
While likelihood analysis is commonly used to draw ecological conclusions, it has been observed that its statistical performance is not robust when faced with slight deviations due to misspecification in SDM.
We propose a new robust estimation method based on a novel divergence for the Poisson point process model.
The proposed method is characterized by weighting the log-likelihood equation to mitigate the impact of heterogeneous observations in the presence-only data, which can result from model misspecification.
We demonstrate that the proposed method improves the predictive performance of the maximum likelihood estimation in our simulation studies and in the analysis of vascular plant data in Japan.  

\


\newpage

\section{Introduction}

Species distribution modeling (SDM) is critical to evaluate biodiversity, conservation and management planning, impacts on human activities, and so on \citep{f13,s18}.
The presence-absence (PA) data for a species of interest require dedicated surveys per spatial unit; thus, such data are difficult and expensive to obtain.
In some cases, the only available data are from museum or herbarium records of locations where the species were observed.
In recent years, the development of geographic information systems has enabled ecologists to obtain knowledge of the environment of the study area without conducting field surveys.
As a result, presence-only (PO) data have become readily available and research based on the PO data has become more active.
However, the PO data often consist of observational surveys rather than the designed surveys, and thus require special attention in the data analysis.

Several statistical and machine learning approaches have been developed to estimate the relative probability or intensity that reflects habitat suitability or abundance of the species of interest.
Some methods are based on the joint likelihood of the presence and environmental variables, or on the partial likelihood \citep{l96,l06b,l09}.
The application of the spatial Poisson point process (PPP) model to the PO data has been proposed \citep{w10,c11}; this model is closely related to the maximum entropy (Maxent) model familiar to ecologists \citep{f13,r13}.
These models can be considered essentially the same method in the sense that they are based on equivalent likelihoods and estimate the equivalent relative probability (or intensity) of presence \citep{w19}.

The present paper focuses on the spatial PPP model.
In the data analysis of the PO data, it is necessary to be aware of bias due to some reasons.
Various methods have been discussed to deal with the sampling bias because the PO data usually suffer from the sampling bias due to heterogeneity in sampling efforts.
These methods involve identifying and estimating the effect of bias on presence and then thinning or eliminating it \citep{d05,f15,k20}.
On the other hand, heterogeneous observations of presence occur due to several reasons, such as incorrect or missing geo-coordinates, taxonomic misidentification, and taxonomic shifts in the PO data \citep{t11,w16,s17}.
In fact, \citet{s17} noted that a median of 10$\%$ (up to 30$\%$) of presence locations per species across 60,065 tree species were either in highly urbanized areas or outside typical habitat areas, which can have a significant effect in assessing habitat suitability.
The heterogeneous observation can have a negative impact on the predictive performance of SDM, e.g., area under the operating characteristic curve (AUC), sensitivity, specificity, and true skill statistic \citep{l18}.
Filtering the dataset improves the performance of SDM; however, the contaminated data cannot be automatically corrected because the specialized knowledge and the time-consuming manual checking are required \citep{b13,m13,v19}.
For large datasets, manually checking the species information introduces exorbitant costs.

Maximum likelihood estimation is usually employed for the PPP model, although it is known that the maximum likelihood estimation is not robust against heterogeneous observations.
Despite the PO data are frequently contaminated by heterogeneous observations, to the best of our knowledge, statistical methods for dealing with this issue are rarely discussed, e.g., an M-estimation \citep{a99}, and a residual analysis for spatial point process \citep{b05}.
Robust methods for parameter estimation have been developed on the basis of information divergences to which the Kullback-Leibler (KL) divergence extends \citep[see, e.g.,][]{b98,f08,e22}.
The present paper proposes a robust parameter estimation based on the Bregman divergence of intensity functions for the thinned PPP models.
The proposed method is characterized by weighting the log-likelihood equation by an influence of heterogeneous observation.

The rest of this paper is organized as follows.
Section \ref{secppp} introduces the spatial PPP model.
Section \ref{secrobust} provides a robust parameter estimation method for the thinned PPP model and discusses the robustness to the heterogeneous observation.
Section \ref{secsim} conducts some simulation studies to evaluate the performance of the proposed method.
Section \ref{secda} presents the analysis of the vascular plant data in Japan.
Finally, Section \ref{secdisc} discusses the findings.

\section{Spatial Poisson point process model}
\label{secppp}

Consider a PPP that occurs in the two-dimensional Euclidean space $\mathbb{R}^2$.
For a study area $\mathscr{A}\subset\mathbb{R}^2$, let $\lambda(s)$ denote an intensity function for any site $s\in\mathscr{A}$ and let $\{s_1,\dots,s_m\}$ denote $m$ presence locations in $\mathscr{A}$.
Assume that (i) the total number $m$ is a sample from a Poisson distribution with intensity $\int_\mathscr{A}\lambda(s)ds$ and that (i\hspace{-.1em}i) the presence locations $(s_i)$ are independent and identically distributed samples of a random variable with the probability function $\lambda(s)/\int_\mathscr{A}\lambda(s)ds$ for $s\in\mathscr{A}$.
The intensity function can be modeled in a complicated form.
For simplicity, this paper assumes a log-linear model for the intensity function; that is, $\log\lambda_{\bmt\beta}(s)=\bm\beta^\top \bm x(s)$, where $\bm\beta=(\beta_0,\beta_1,\dots,\beta_p)^\top$ is a coefficient parameter vector, and $\bm x(s)=(1,x_1(s),\dots,x_p(s))^\top$ is an environmental variable vector at a location $s$.
To deal with the bias due to the heterogeneity of sampling effort, the thinned PPP was developed \citep{d05,f15}.
Consider the modeling of a PPP by the thinning PPP with the intensity $\lambda_{\bmt\theta}(s)=\lambda_{\bmt\beta}(s)b_{\bmt\alpha}(s)$ using a detection probability $b_{\bmt\alpha}(s)$ for a site $s\in\mathscr{A}$, where $\bm\alpha=(\alpha_1,\dots,\alpha_q)^\top$ is a parameter vector and $\bm\theta=(\bm\beta^\top,\bm\alpha^\top)^\top$.
This paper assumes a logistic-linear model for the detection probability; that is, $\logit b_{\bmt\alpha}(s)=\bm\alpha^\top \bm z(s)$, where $\bm z(s)=(z_1(s),\dots,z_q(s))^\top$ is a vector of sampling-bias variable, e.g., distance from a road or an urbanized area.

The log-likelihood function of the thinned PPP model is defined by
\begin{eqnarray}
  l(\bm\theta) = \sum_{i=1}^m\log\left\{\lambda_{\bmt\theta}(s_i)\right\} - \int_\mathscr{A}\lambda_{\bmt\theta}(s)ds. \label{like}
\end{eqnarray}
Noting that the integral over a study area in equation (\ref{like}) cannot be exactly calculated, a numerical approximation method has been proposed to estimate the integral using quadrature weights \citep{b92}.
Without loss of generality, we set up a location vector $\{s_1,\dots,s_r\}$ when the study area $\mathscr{A}$ is split into $n$ grid cells, where $r=n+m-m^{(n)}$, $m^{(n)}$ is the number of grid cells that contain at least one presence location, and $\{s_{m+1},\dots,s_r\}$ are the centers of the grid cells that contain no presence location.
The approximated log-likelihood function is then given by
\begin{eqnarray}
  l(\bm\theta) = \sum_{i=1}^r\left[ d_i\log\left\{\lambda_{\bmt\theta}(s_i)\right\}-w_i\lambda_{\bmt\theta}(s_i) \right], 
\end{eqnarray}
where $d_i=I(i\in\{ 1,\dots,m \})$, $I(\cdot)$ is the indicator function, and $w_i$ is a quadrature weight for a location $s_i$.
We regard the area of a grid cell divided by the number of locations $\{ s_1,\dots,s_r \}$ contained in the cell as the quadrature weight $w_i$ in the same manner as \citet{r13}.
That is, $w_i=|\mathscr{A}|/(nm_i^+)$, where $|\mathscr{A}|$ is the area of $\mathscr{A}$, $m_i^+=\max\{1,m_i\}$, and $m_i$ is the number of presence locations at a grid cell $s_i$. 
The likelihood equations can be obtained by
\begin{align}
  \frac{\partial}{\partial\bm\beta}l(\bm\theta) & = \sum_{i=1}^r\varepsilon_{\bmt\beta}(\bm\theta,s_i) = \sum_{i=1}^r\{ d_i - w_i\lambda_{\bmt\theta}(s_i) \}\bm x(s_i)=\bm 0_{1+p}, \label{eemle} \\
  \frac{\partial}{\partial\bm\alpha}l(\bm\theta) & = \sum_{i=1}^r\varepsilon_{\bmt\alpha}(\bm\theta,s_i) = \sum_{i=1}^r\frac1{1+\exp\left(\bm\alpha^\top \bm z(s_i)\right)}\{ d_i - w_i\lambda_{\bmt\theta}(s_i) \}\bm z(s_i)=\bm 0_q, \label{eemle2}
\end{align}
where $\bm 0_k$ indicates a $k$-dimensional zero vector.
Solving the likelihood equations, the maximum likelihood estimator (MLE) of parameter $\bm\theta$ is obtained.
The intensity function is estimated by plugging the MLEs into the intensity model, $\lambda_{\bmt\beta}(s)=\exp\{\bm\beta^\top\bm x(s)\}$, and setting $\bm\alpha=\bm 0$ that means the sampling-bias effect is removed.

If the species distribution model is correctly specified, then the maximum likelihood method would yield accurate conclusions, even in the presence of sampling bias.
However, we must be cautious in situations where model misspecification occurs in practical ecological studies.
A non-negligible degree of misspecification can render the MLE unreliable and lead to incorrect inference.
For instance, unobservable feature variables might cause the underlying intensity function to deviate slightly from parametric intensity functions, such as interactions with other species. 
The following section will address this issue and propose a new estimation method.

\section{Robust parameter estimation}
\label{secrobust}

\begin{figure}[t]
\begin{center}
\includegraphics[width=12cm]{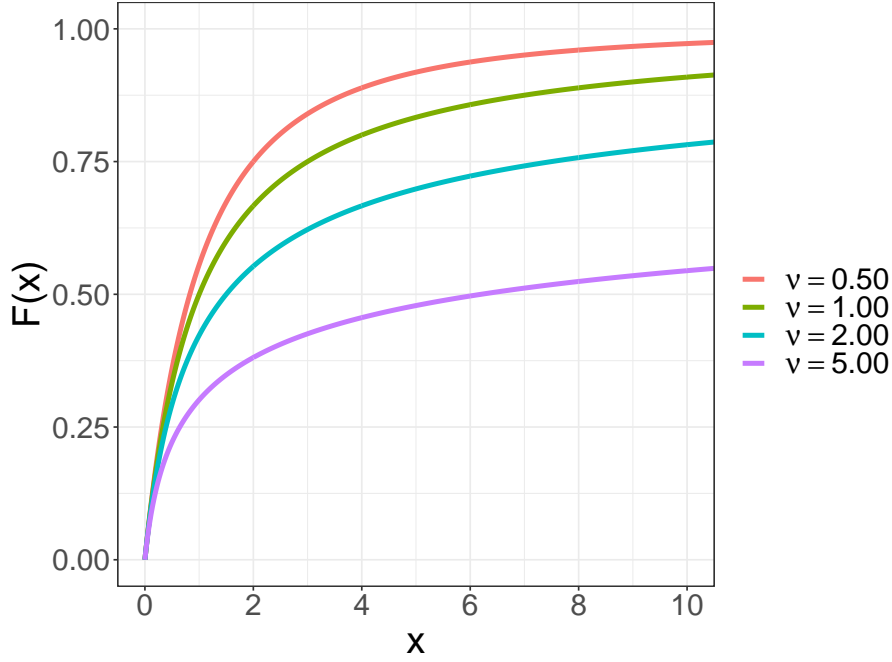}\\
\caption{Plots of the Parato type II distribution function with $\nu=0.5, 1.0, 2.0, 5.0$.}
\label{figcdf}
\end{center}
\end{figure}

We are concerned with various possibilities for the misspecification of the thinned PPP model with the parametric intensity function $\lambda_{\bmt\theta}(s)$ as introduced in section \ref{secppp}.
Our main objective is to propose a robust estimation method for the parameter $\bm\theta$.
To achieve this, we employ a weighted likelihood equation approach. 
We focus on the values of  parametric intensities $\{\lambda_{\bmt\theta}(s_i): i=1,..,r\}$, which should represent  the species abundance.
Suppose that occurrences are partially generated by a heterogeneous point process different from the assumed model with the parameter $\bm\theta$. 
In this case, the values of parametric intensities corresponding to heterogeneous processes have a small magnitude.
In light of this, we propose a weighted likelihood equation for $\bm\theta=(\bm\beta^\top,\bm\alpha^\top)^\top$ as follows:
\begin{align}
\sum_{i=1}^rF(\tau\lambda_{\bmt\theta}(s_i))\varepsilon_{\bmt\beta}(\bm\theta,s_i) &= \bm 0_{1+p}, \label{eeb} \\
\sum_{i=1}^rF(\tau\lambda_{\bmt\theta}(s_i))\varepsilon_{\bmt\alpha}(\bm\theta,s_i) &= \bm 0_q, \label{eeb2}
\end{align}
where $\tau > 0$ is a constant tuning parameter and $F(\cdot)$ is a cumulative distribution function for the value of the intensity function.
We refer to the estimator of $\bm\beta$ of interest, say $\hat{\bm\beta}_\tau$, based on equations (\ref{eeb}) and (\ref{eeb2}) as the minimum intensity divergence estimator (MIDE).
Here we adopt the Pareto type II distribution
\begin{eqnarray}
F(x)=1-\big(1+{\nu}x\big)^{-\frac{1}{\nu}}
\end{eqnarray}
for $x>0$, where $\nu>0$ is a shape parameter.
See Figure \ref{figcdf}. 
This distribution has a long tail, which achieves stable estimation by preventing extremely large weights.
For practical purposes, we will fix $\nu=1$.
The weight function $F(\tau \lambda(s_i,\theta))$ expresses the magnitude of the intensity function at point $s_i$, and calibrates the model correctness by suppressing the influence of heterogeneous observations in the weighted likelihood function.
We derive the estimating equations  (\ref{eeb}) and (\ref{eeb2}) in terms of minimization of a specific example of Bregman divergence from the true intensity function to the parametric intensity function $\lambda_{\bmt\theta}(s)$.
A detailed derivation of equations (\ref{eeb}) and (\ref{eeb2}) are provided in \ref{appmide}.
The property of the divergence automatically leads to the consistency of the proposed estimator for $\bm\theta$.
The consistency and asymptotic normality of the MIDE can be derived (see \ref{appmide2} for the details).
When the tuning parameter $\tau$ goes to $\infty$, then the MIDE reduces to the MLE from equations (\ref{eemle}) and (\ref{eemle2}) because all the weights of the estimating equations are equal to 1.
Therefore, the MLE has no chance of suppressing the influence of heterogeneous observations.

To simultaneously carry out the shrinkage estimation and the variable selection, we consider the penalized loss function with an $L_1$ penalty \citep{t96} as follows.
\begin{eqnarray}
l^\phi_\Xi(\bm\beta,\bm\alpha) = l_\Xi(\bm\beta,\bm\alpha) + \sum_{k=0}^p\phi_k|\beta_k|, \label{lossp}
\end{eqnarray}
where $l_\Xi$ is the loss function minimized by solving the estimating equations (\ref{eeb}) and (\ref{eeb2}) and is given by equation (\ref{lossb}) in \ref{appmide}, $\phi_0=0$, and $\phi_k=\phi$ ($k\neq 0$, $\phi\geq 0$) is a constant tuning parameter.
The loss function has no penalties for the intercept parameter $\beta_0$ of the intensity function and the coefficient paramter $\bm\alpha$ for the detection probability model.
The gradient ascent method \citep{g10} can be used to compute the $L_1$ penalized estimates.
The detailed computation algorithm is provided in \ref{appga}.

The root trimmed mean squared prediction error (RTMSPE) is employed to select the appropriate values of the tuning parameters $\tau$ and $\phi$ by removing heterogeneous observations.
The RTMSPE is given by
\begin{eqnarray}
  {\rm RTMSPE}_\delta = \sqrt{\frac1{h_\delta}\sum_{i=1}^{h_\delta}e^2_{[i]}}\qquad(0<\delta<1),
\end{eqnarray}
where $h_\delta=\lfloor (n+1)\delta \rfloor$ and $e^2_{[1]}\leq\cdots\leq e^2_{[n]}$ are the order statistics of $[m_{[1]}-\lambda_{\hat{\bmt\theta}}(s_{[1]})]^2,\dots,[m_{[n]}-\lambda_{\hat{\bmt\theta}}(s_{[n]})]^2$ for grid cells ($s_{[1]},\dots,s_{[n]}$) that do not overlap each other and the number of observations ($m_{[1]},\dots,m_{[n]}$) in the cells with a estimate $\hat{\bm\theta}$.

\section{Simulation}
\label{secsim}

We evaluated the robustness of the MIDE compared with that of the MLE.
Two types of situations were considered: one where the data were generated from the target PPP distribution with no contamination, and another where the data were contaminated by heterogeneous distribution.

Consider target and contamination PPPs on a study area $\mathscr{S}$ divided into 2,000 grid cells.
The intensity function of the target PPP is $\lambda_{\bmt\beta}(s)=\exp(\bm\beta^\top \bm x(s))$, $\bm\beta=(\beta_{0},\beta_{1},\beta_{2},\beta_{3},\beta_{4})^\top$, and that of the contamination PPP is $\lambda_{\bmt\gamma}(s)=\exp(\bm\gamma^\top \bm x(s))$, $\bm\gamma=(\gamma_{0},\gamma_{1},\gamma_{2},\gamma_{3},\gamma_{4})^\top$.
We note that $\lambda_{\bmt\gamma}(s)$ is in the log-linear model, but the parameter $\bm\gamma$ is specified a totally different value of $\bm\beta$.
The detection probability is $b(s)=\expit(\bm\alpha^\top \bm z(s))$, $\bm\alpha=(1,-1)^\top$.
In the no contamination case, the simulated data were sampled from the thinned target distribution with the intensity $\lambda_{\bmt\beta}(s)b(s)$.
In the contamination case, the data were sampled from a thinned superposed PPP for the target and contamination distributions with the intensity $(\lambda_{\bmt\beta}(s)+\lambda_{\bmt\gamma}(s))b(s)$.
The expected contamination rate is $\Sigma_s\lambda_{\bmt\gamma}(s)/(\Sigma_s\lambda_{\bmt\beta}(s)+\Sigma_s\lambda_{\bmt\gamma}(s))$.
The true values of parameters $\bm\beta$ and $\bm\gamma$ were set such that the expected contamination rate was sufficiently small.
Moreover, the intensity of the contamination distribution was relatively large in some areas where the intensity of the target distribution was relatively small.
Therefore, we set $\gamma_i=-\beta_i$, $i\neq 0$.
The environmental variable and the bias variable were generated from the standard normal distribution.
The number of presence locations $m$ was generated from a Poisson distribution with the mean which is the total intensity summed over the study area.
The presence locations were then generated from the multinomial distribution with the probability which is the intensity divided by the total intensity.

We investigated the performances of MLE and MIDE via three simulation scenarios: a no-contamination case, and light- and heavy-contamination cases with expected contamination rates of about 10$\%$ and 20$\%$, respectively.
The 200 datasets were simulated in each case.
The tuning parameter $\tau$ was selected by a grid search among the candidate values $\{0.1,1,5,10,20,\infty\}$ to minimize the RTMSPE$_{0.9}$ in simulations.
The details of the simulation settings and results for the slope parameters of interest are provided in Figures \ref{figsim}a-c.
In addition, the selected percentages of values of tuning parameter $\tau$ are given in Table \ref{tabsim}.
In the no-contamination case, both the MLE and MIDE correctly estimated the true values of the coefficient parameters.
For the MIDE, 21.5$\%$ of simulations selected $\tau$ as $\infty$ (the MLE).
In both the light- and heavy-contamination cases, the MIDEs of the coefficient parameters were close to the true values of the target distribution, whereas the MLEs were not.
For the results of MIDE, the variations of estimates were slightly larger in the heavy-contamination case than in the light-contamination case; however, the estimates were given around the true values in both cases.
The value of $\tau$ was selected as not $\infty$ in 96.5-100$\%$ of the simulations, and the MLE was selected as 0-3.5$\%$ in the light- and heavy-contamination cases.

\begin{figure}[H]
\begin{minipage}{1\hsize}
\begin{center}
\includegraphics[width=9cm]{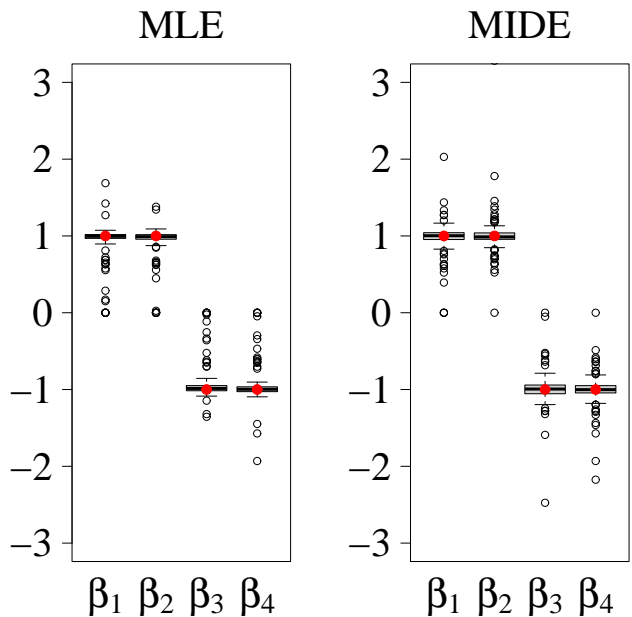} \qquad \qquad

(a) No-contamination case
\end{center}
\end{minipage}
\caption{Boxplots of the parameter estimates of the MLE and MIDE for
(a) $\bm\beta=(-2, 1, 1, -1, -1)^\top$ in the no-contamination case,
(b) $\bm\beta=(-2, 1, 1, -1, -1)^\top$ and $\bm\gamma=(-4.2, -1, -1, 1, 1)^\top$ in the light-contamination case,
(c) $\bm\beta=(-2, 1, 1, -1, -1)^\top$ and $\bm\gamma=(-3.4, -1, -1, 1, 1)^\top$ in the heavy-contamination case.
Red points indicate the true values.
The tuning parameters $\tau$ and $\phi$ were selected on the basis of the RTMSPE$_{0.9}$.}
\label{figsim}
\end{figure}

\addtocounter{figure}{-1}

\begin{figure}[H]
\begin{minipage}{1\hsize}
\begin{center}
\includegraphics[width=9cm]{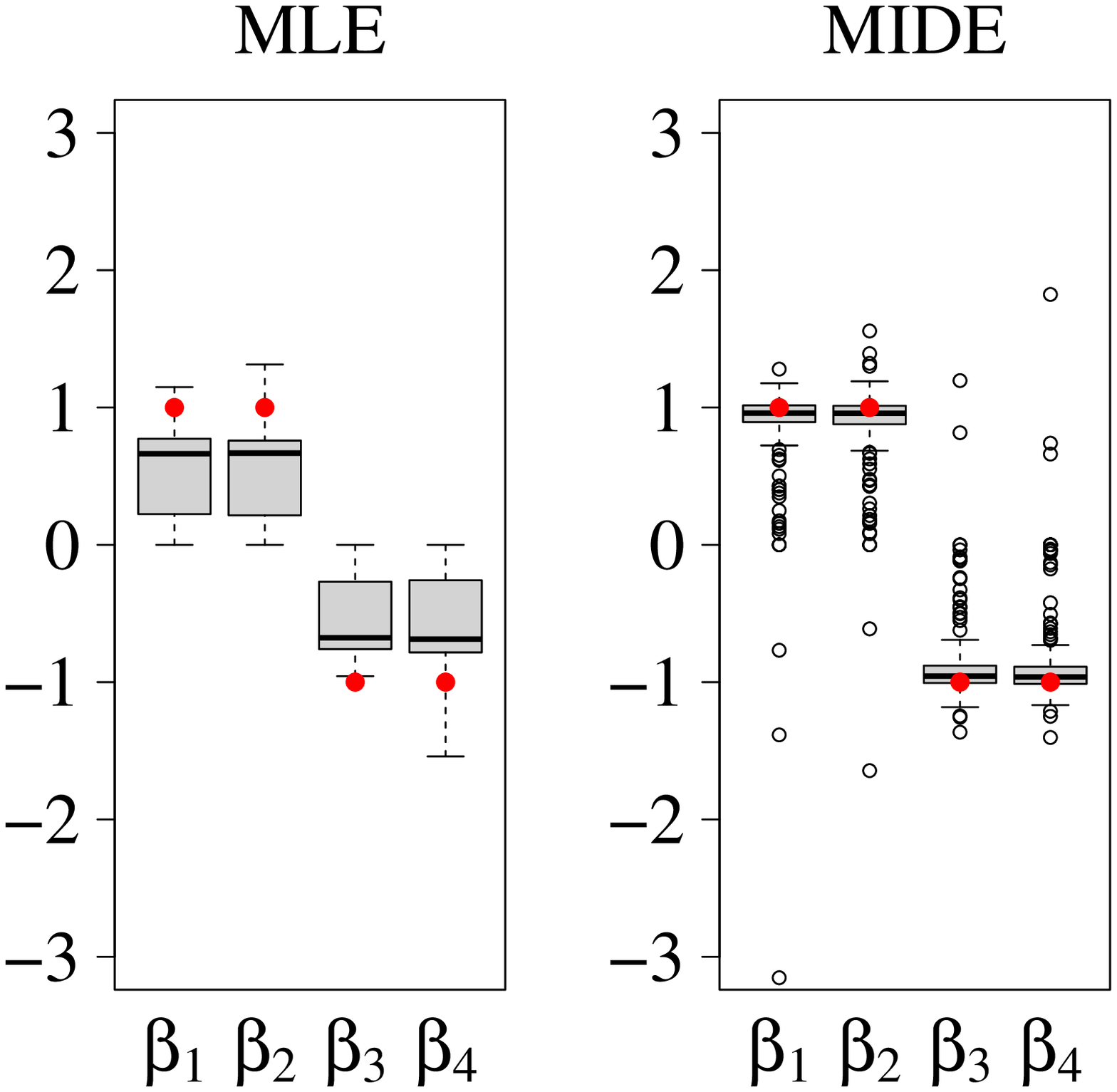} \qquad \qquad

(b) Light-contamination case
\end{center}
\end{minipage}

\vspace{1em}

\begin{minipage}{1\hsize}
\begin{center}
\includegraphics[width=9cm]{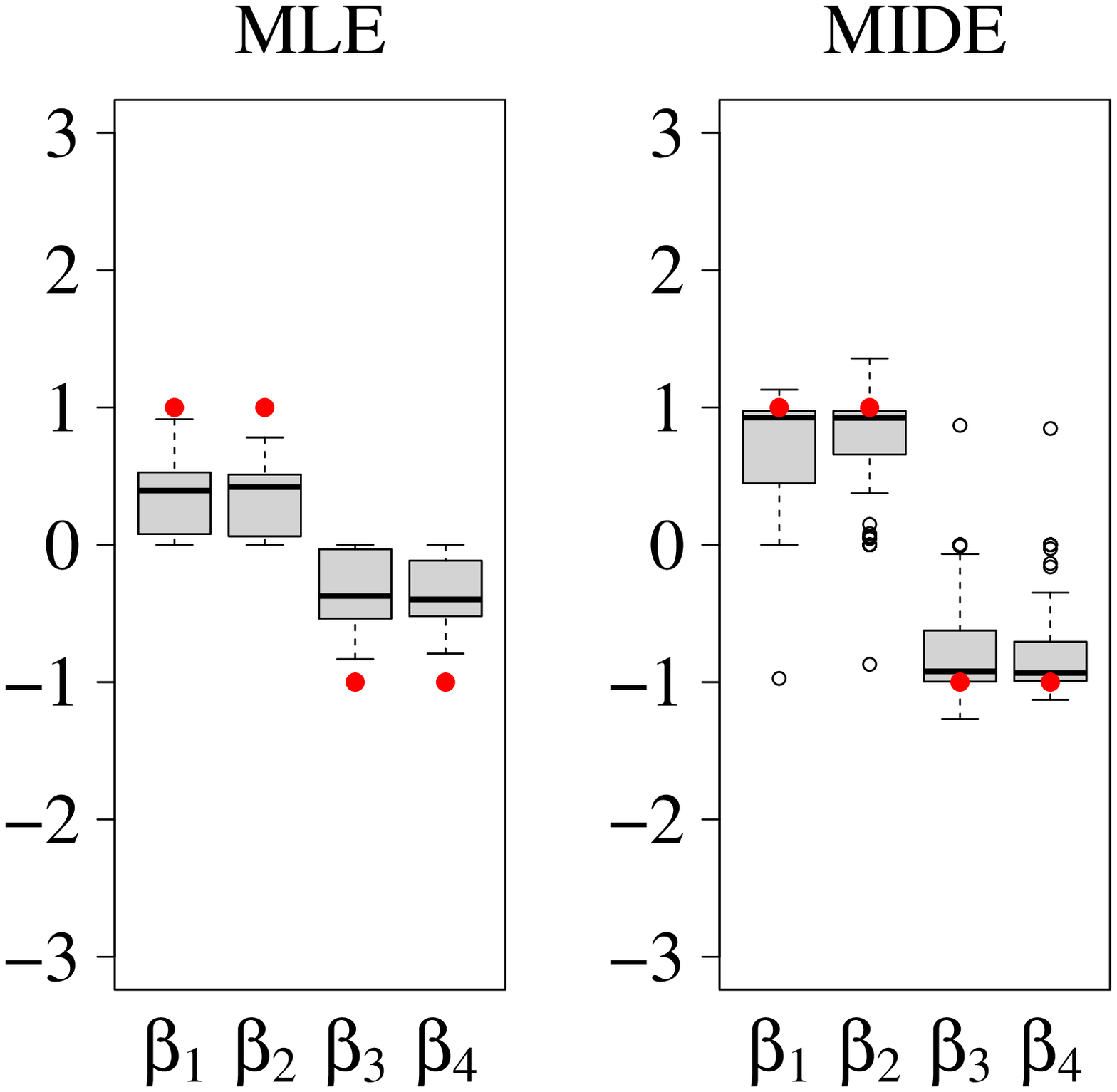}

(c) Heavy-contamination case
\end{center}
\end{minipage}
\caption{(continued)}
\end{figure}

\begin{table}[t]
\caption{Percentages of simulations when the tuning parameter $\tau$ of the MIDE was selected to be each value in the simulations with no-, light- and heavy-contamination cases for $q$ of the RTMSPE.}
\begin{center}
\noindent
\begin{tabular}{cccccc}
\hline
\multicolumn{6}{c}{ Value of $\tau$ of the MIDE}\tabularnewline
\cline{1-6}
\multicolumn{1}{c}{0.1}&\multicolumn{1}{c}{1}&\multicolumn{1}{c}{5}&\multicolumn{1}{c}{10}&\multicolumn{1}{c}{20}&\multicolumn{1}{c}{$\infty$ (MLE)}\tabularnewline
\hline
\multicolumn{6}{l}{(a) No-contamination case}\\
0.063&0.038&0.025&0.253&0.405&0.215\tabularnewline  \\
\multicolumn{6}{l}{(b) Light-contamination case}\\
0.050&0.110&0.100&0.245&0.460&0.035\tabularnewline  \\
\multicolumn{6}{l}{(c) Heavy-contamination case}\\
0.030&0.134&0.164&0.373&0.299&0.000\tabularnewline
\hline
\end{tabular}\end{center}
\label{tabsim}
\end{table}

\section{Data analysis}
\label{secda}

The PO data and the independent PA data of vascular plants in Japan have been compiled in \citet{k15b}.
These two datasets for 40 species are available in the R package `qPPP' \citep{k20}.
The study area was split into 10km $\times$ 10km grid cells ($n=4,684$) on the basis of a regular mesh in the PO data.
Duplicate presence locations in each cell were removed to reduce spatial clumping and to avoid inflating of the model accuracy \citep{v09}.
The study area was divided into seven subregions: Hokkaido, Tohoku, Kanto, Chubu, Kinki, Chugoku-Shikoku, and Kyushu regions.
Subregions with a small number of observations or large discrepancy between the observations of PO and PA data were excluded from the analysis.
Therefore, the thinned PPP model was applied to the 27 combinations of species and subregions where $m\geq 50$ and the proportion of presence locations of intersection of the PO and PA data to those of the union was greater than 0.5. 
We then calculated the MLE and MIDE with the tuning parameter $\tau$ selected among $\{0.1,1,5,10,20,\infty\}$ to minimize the RTMSPE$_{0.9}$.
The 37 environmental variables included in the qPPP package were used as shown later.
The sampling-bias variable was the number of presence in the target group \citep{d05,p09}.
All the environmental and bias variables were standardized to have mean 0 and variance 1.
The predictive performances of the MLE and MIDE were evaluated on the basis of the AUC calculated from the independent PA data.

\begin{figure}[H]
\begin{minipage}{1\hsize}
\begin{center}
\includegraphics[width=9cm]{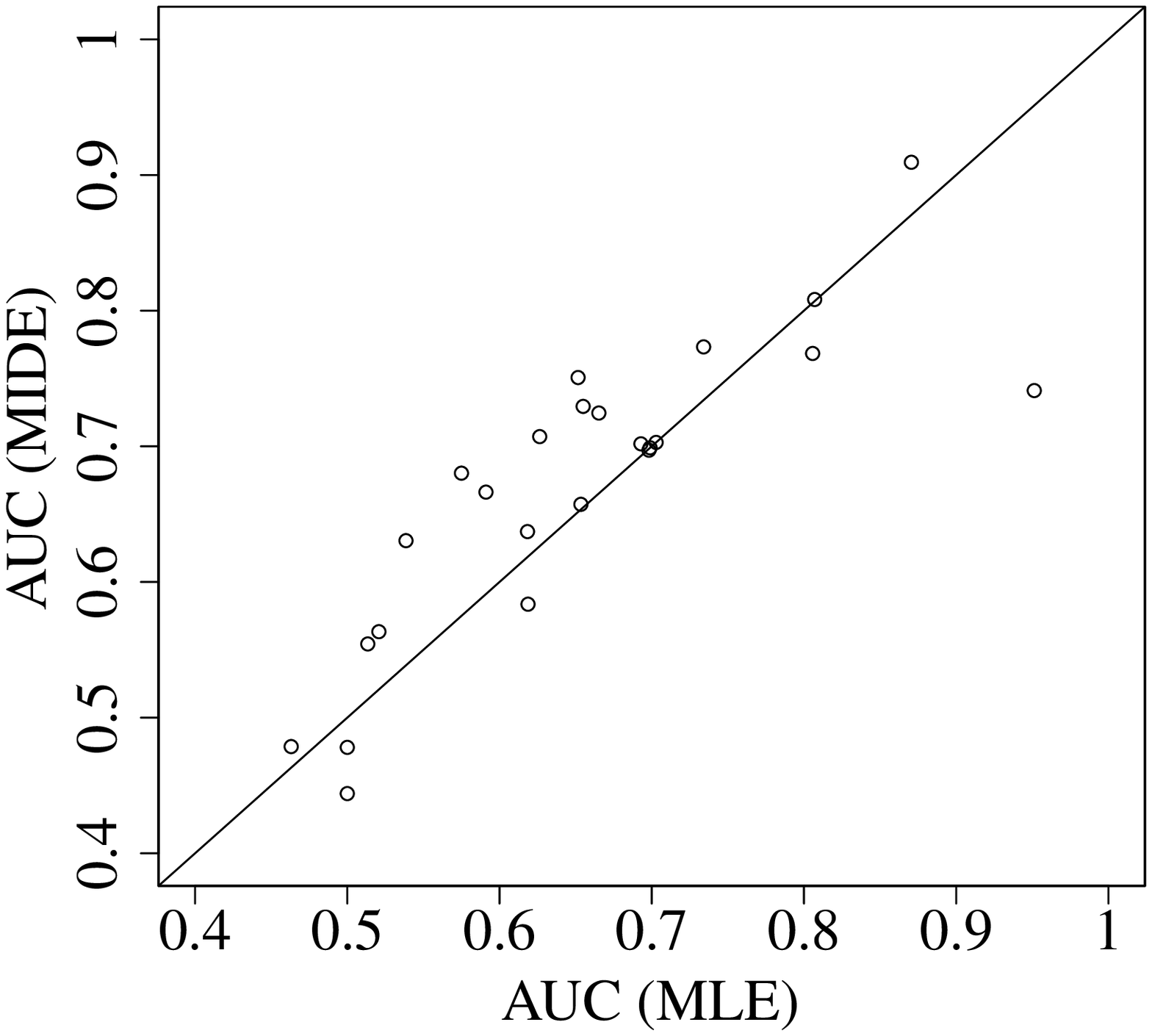} 
\end{center}
\end{minipage}
\caption{AUCs calculated from the PA data on the basis of the MLE and MIDE for combinations of species and subregions.}
\label{figauc}
\end{figure}

Fig \ref{figauc} displays the AUCs calculated from the PA data based on the MLE and MIDE.
This figure shows that the AUCs for the MIDE were similar to or improved over those for the MLE in most combinations of species and subregions.

\begin{figure}[H]
\begin{minipage}{0.5\hsize}
\begin{center}
\includegraphics[width=8cm]{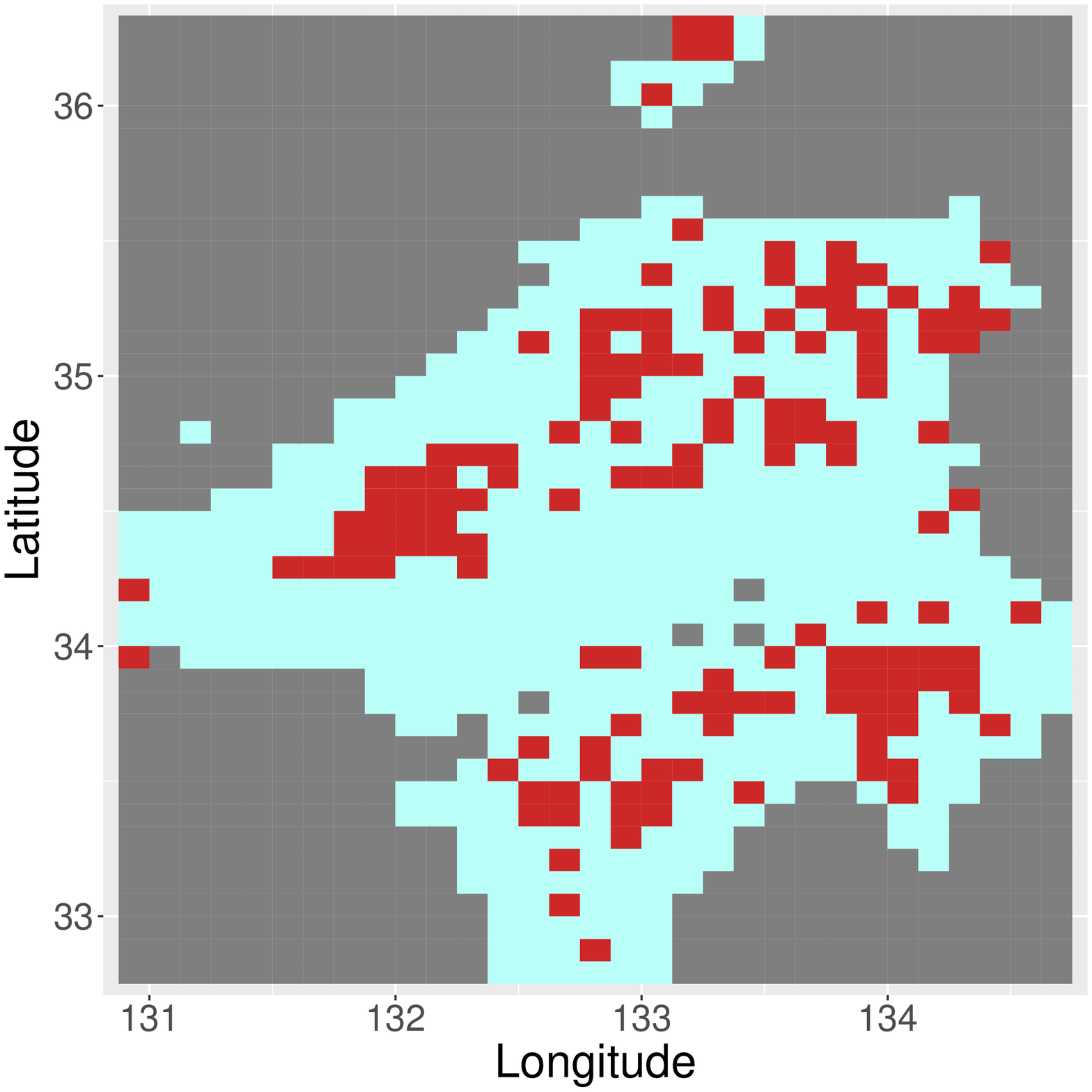} \qquad \qquad

(a) PO data
\end{center}
\end{minipage}
\begin{minipage}{0.5\hsize}
\begin{center}
\includegraphics[width=8cm]{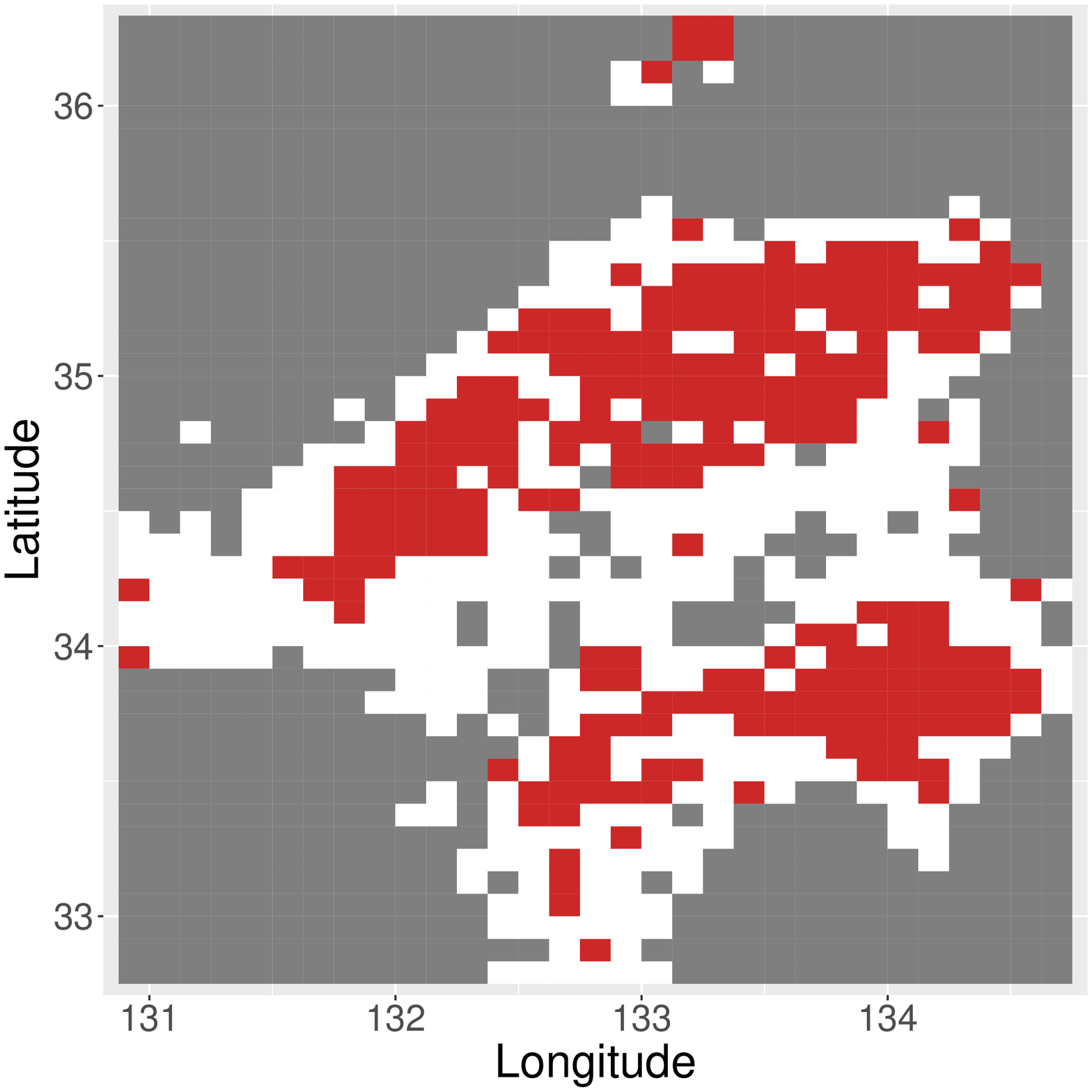} \qquad \qquad

(b) PA data
\end{center}
\end{minipage}

\vspace{1em}

\begin{minipage}{0.5\hsize}
\begin{center}
\includegraphics[width=8cm]{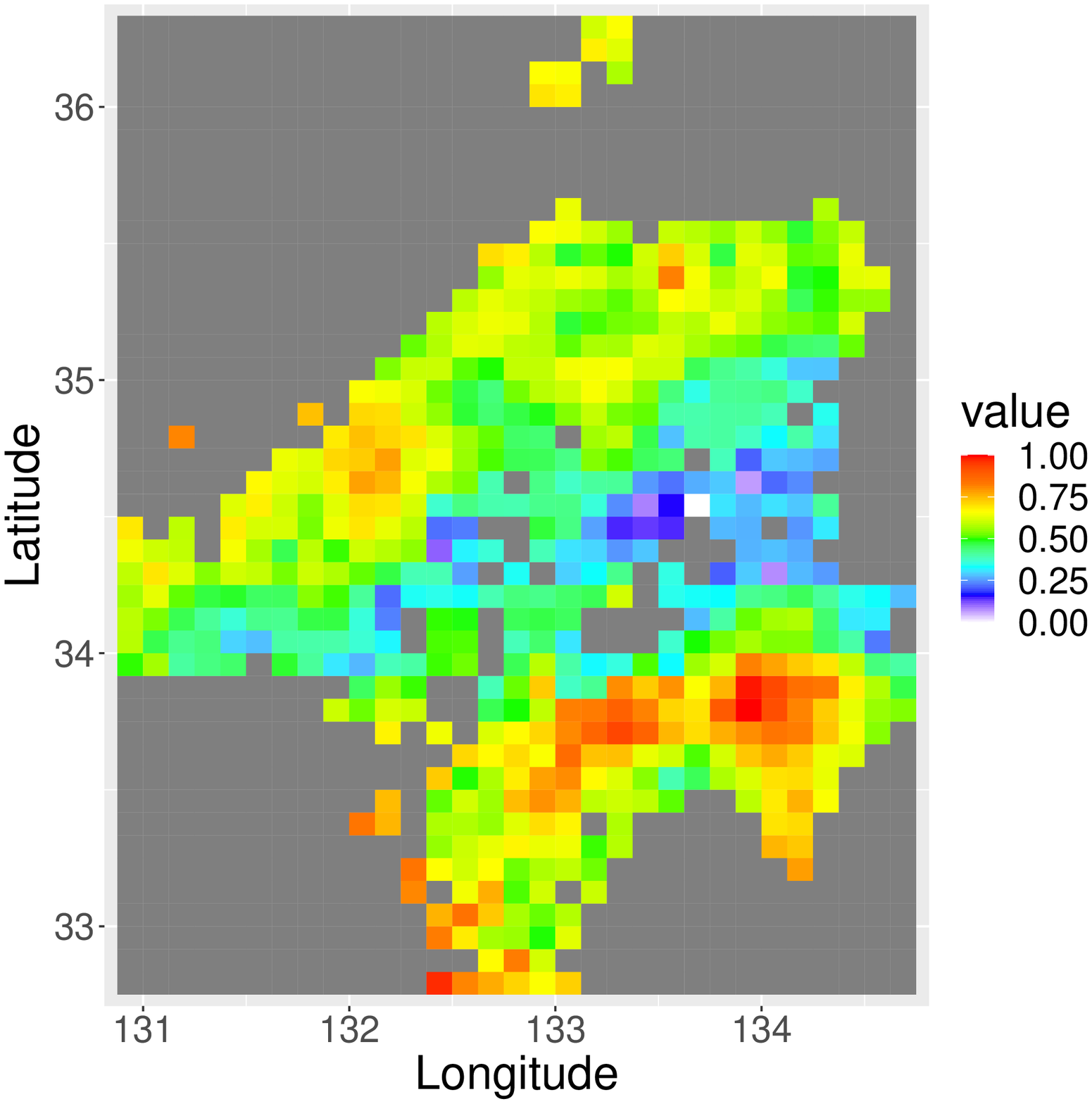}

(c) MLE
\end{center}
\end{minipage}
\begin{minipage}{0.5\hsize}
\begin{center}
\includegraphics[width=8cm]{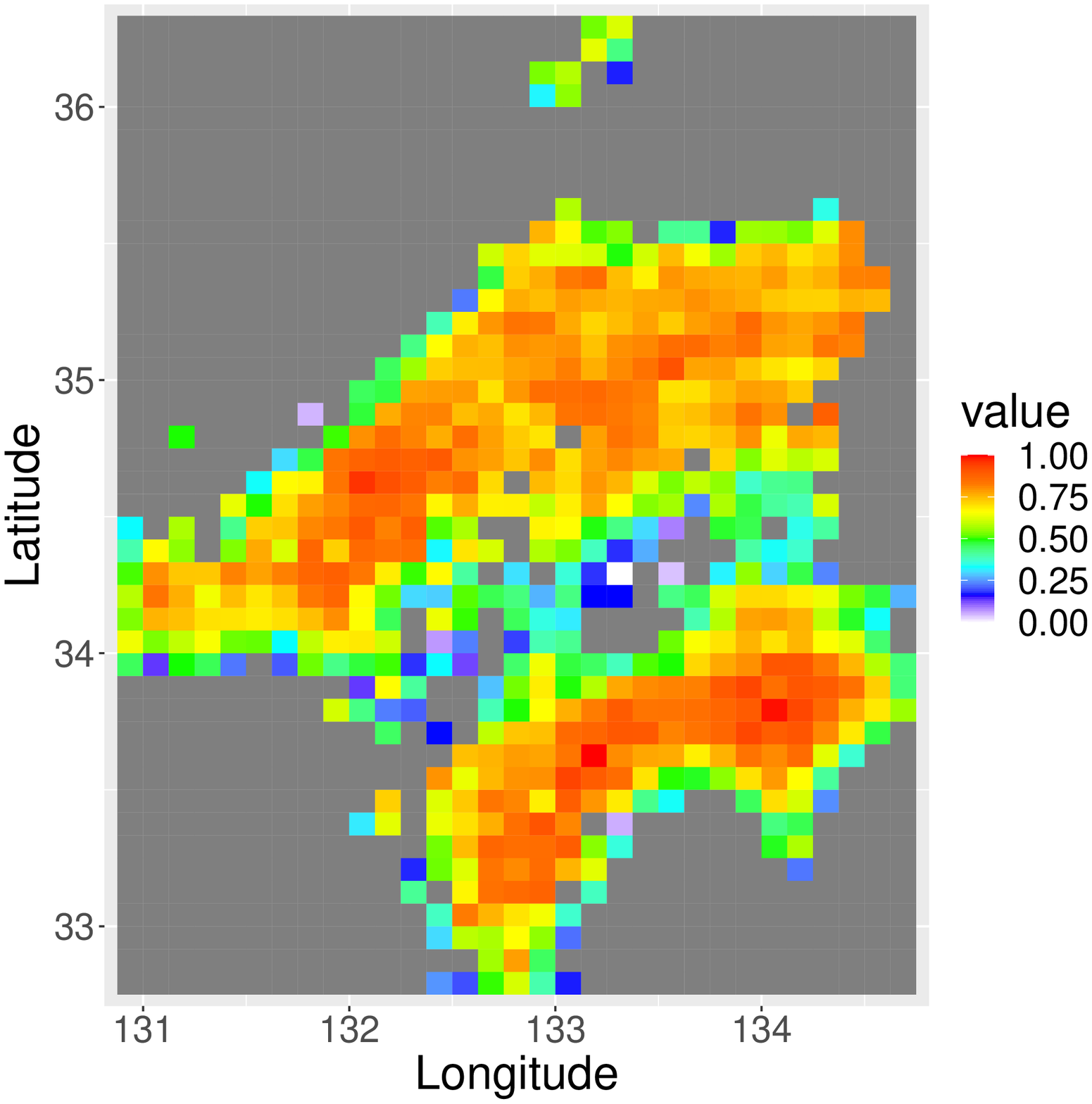}

(d) MIDE
\end{center}
\end{minipage}
\caption{The presence locations of $Carpinus$ $Laxiflora$ in the Chugoku-Shikoku region (a) in the PO data and (b) in the PA data. 
The red tiles indicate the presence locations.
The light blue tiles in (a) indicate the pseudo-absence locations, and the white tiles in (b) indicate the absence locations.
The standardized predicted intensity functions of the (c) MLE and (d) MIDE, where higher values indicate better habitat suitability.}
\label{figmap}
\end{figure}

\begin{figure}[t]
\begin{center}
\includegraphics[height=10cm]{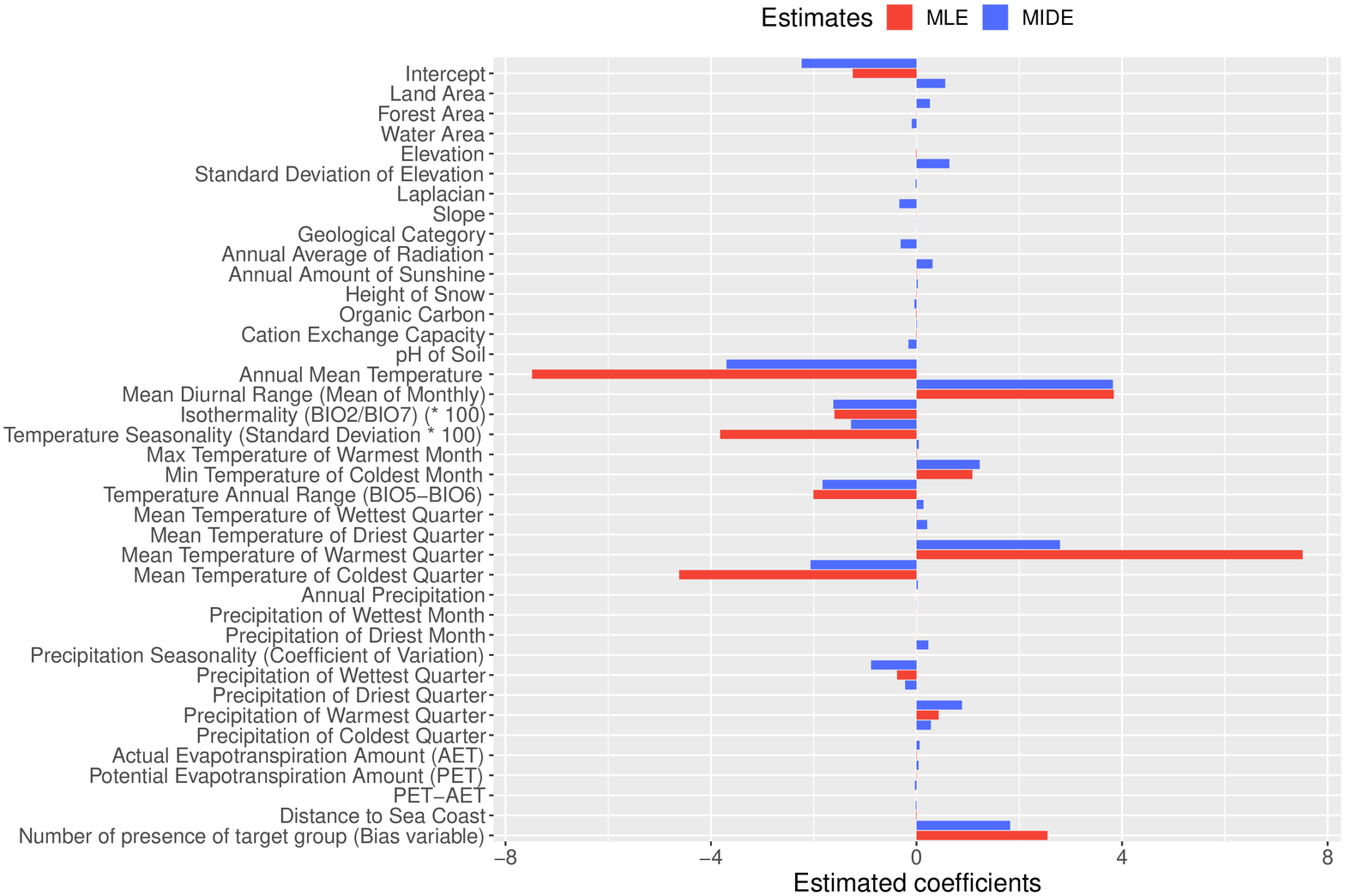}\\
\caption{The MLE and MIDE of the regression parameter $\bm\beta$ for the environmental and bias variables for $Carpinus$ $Laxiflora$ in the Chugoku-Shikoku region.}
\label{figcoef}
\end{center}
\end{figure}

\begin{figure}[t]
\begin{center}
\includegraphics[width=12cm]{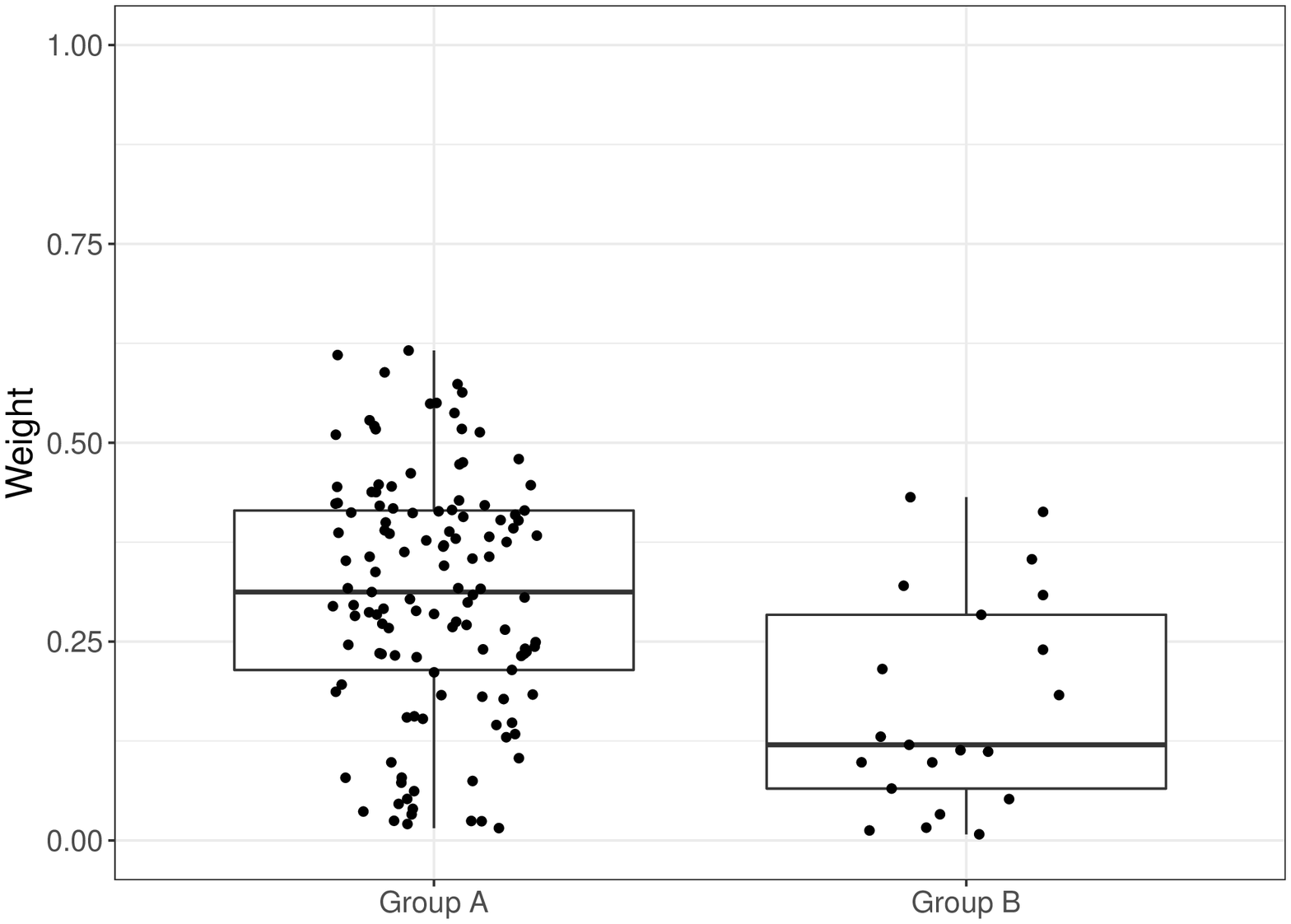}\\
\caption{Boxplots of the weight of the estimating function of the MIDE for two groups. 
Group A is the presence locations in the PO data that have the presence locations in the PA data within the area of a disc with a radius of 5km centered at the location, whereas Group B is the other presence locations in the PO data.}
\label{figweight}
\end{center}
\end{figure}

The PO and PA data of $Carpinus$ $Laxiflora$ in the Chugoku-Shikoku region are illustrated in Figures \ref{figmap}a-b.
The number of presence locations is 121, and the number of pseudo-absence locations is 254 in the PO data.
Most of the presence locations in the PA data are included among the presence locations in the PO data.
Conversely, if the PA data are considered as the true distribution, some presence locations included only in the PO data may be suspicious.
The predictive performance was improved; the AUC based on the MLE was $0.626$, and that based on the MIDE with the selected value $\tau=1$ was $0.707$.
The standardized predicted intensity functions for the PA data of the MLE and MIDE are displayed in Figures \ref{figmap}c-d.
Some characteristic differences between the predicted intensity functions based on the MLE and MIDE were observed.
In particular, results based on MIDE showed that the northern region has a relatively high habitat suitability, whereas results based on MLE showed no such trend.
The estimated regression coefficients for the environmental variables were consistent in sign between the MLE and MIDE, although differences were observed in the magnitude of the values (Figure \ref{figcoef}).
The estimated coefficient on the bias variable was positive and may correctly reflect the amount of sampling effort.

To explain the improvement in the AUC, we compared the weights of the estimating function of the MIDE between two groups of the presence locations in the PO data.
One group had presence locations in the PA data within the area of a disc of radius of 5km centered at the location and the other group had no presence location in the vicinity.
That is, the former locations were relatively reliable, whereas the latter locations might be associated with suspicious data.
The weights for the latter locations were relatively smaller than those for the former locations (Figure \ref{figweight}).
Therefore, the MIDE might reduce the effect of suspect observational information by adding weights to the estimation function, resulting in better prediction results.

\section{Discussion}
\label{secdisc}

We proposed a new robust estimation method for the thinned PPP model against heterogeneous observations in which a species is observed even though the probability (or intensity) of its presence is low.
The weight of the estimating function of the proposed estimator plays an important role in the robust and stable estimation.

We attempted to consider estimators based on information divergences other than the proposed divergence, but they were unstable. 
The loss functions can be considered based on the $\beta$- and $\gamma$-divergences \citep{b98,f08}.
Then the estimating functions have weights that increases exponentially with the magnitude of the model intensity.
In our simulation studies, the results based on the estimating functions differed dramatically and the computation algorithm did not converge when the value of the tuning parameter which is an exponent was slightly changed.
At that time, the weights of the power of the intensity function were extremely large for a small subset of the data, which might have led to the unstable estimation.
On the other hand, the proposed estimator uses a weight that is rescaled by a distribution function $F$ and drops the weight into the range from 0 to 1, which might have achieved the stable estimation.

The proposed method can be used in combination with some other useful statistical methods.
The methods to reduce sampling bias by filtering the dataset (e.g., the target-group background method \citep{p09}, and the spatial thinning \citep{v09} employed in Section \ref{secda}) can be used simultaneously.
If there is spatial dependence (or autocorrelation) in species distribution \citep{d07}, the proposed estimator can be used to train a spatial dependence model such as the area-interaction process \citep{b95}.
Further study is needed for robust estimation by the proposed method of mixed-effects models such as the Cox process \citep{m98}.


\

\noindent
{\large \bf Author Contributions}

SE conceived the ideas and designed methodology; YS and OK conducted the simulation and analysed the data; YS wrote the original draft; All authors contributed critically to the drafts and gave final approval for publication.

\

\noindent
{\large \bf Supplemental Material}

An R code implementing the proposed method is available at GitHub\\ \noindent(https://github.com/saigusay/MIDE).

\

\noindent
{\large \bf References}

\begin{description}

\bibitem[\protect\citeauthoryear{Assun\c{c}\~{a}o and Guttorp}{1999}]{a99}
Assun\c{c}\~{a}o R, Guttorp P. (1999).
Robustness for inhomogeneous Poisson point processes. 
Annals of the Institute of Statistical Mathematics 51(4), 657-678.

\bibitem[\protect\citeauthoryear{Baddeley and van Lieshout}{1995}]{b95}
Baddeley AJ, van Lieshout MNM. (1995). 
Area-interaction point processes.
Annals of the Institute of Statistical Mathematics 47(4), 601–619.

\bibitem[\protect\citeauthoryear{Baddeley et~al.}{2005}]{b05}
Baddeley AJ, Turner R, M{\o}ller J, Hazelton M (2005).
Residual analysis for spatial point processes. 
Journal of the Royal Statistics Society, Series B 67, 617-666.

\bibitem[\protect\citeauthoryear{Basu~et~al.}{1998}]{b98}
Basu A, Harris IR, Hjort NL, Jones MC. (1998).
Robust and efficient estimation by minimising a density power divergence.
Biometrika 85(3), 549-559.

\bibitem[\protect\citeauthoryear{Belbin~et~al.}{2013}]{b13}
Belbin L, Daly J, Hirsch T, Hobern D, La Salle J. (2013). 
A specialist\mbox{’}s audit of aggregated occurrence records: An \mbox{`}aggregator\mbox{’}s\mbox{’} perspective.
ZooKeys 305, 67-76. 

\bibitem[\protect\citeauthoryear{Berman and Turner}{1992}]{b92}
Berman M, Turner T. (1992). 
Approximating point process likelihoods with GLIM. 
Journal of the Royal Statistics Society, Series C 41(1), 31-38.

\bibitem[\protect\citeauthoryear{Chakraborty~et~al.}{2011}]{c11}
Chakraborty A, Gelfand AE, Wilson AM, Latimer AM, Silander JA. (2011).
Point pattern modelling for degraded presence-only data over large regions. 
Journal of the Royal Statistical Society, Series C 60(5), 757-776.

\bibitem[\protect\citeauthoryear{Dormann}{2007}]{d07}
Dormann CF. (2007). 
Effects of incorporating spatial autocorrelation into the analysis of species distribution data. 
Global Ecology and Biogeography 16(2), 129–138.

\bibitem[\protect\citeauthoryear{Dud\'{i}k~et~al.}{2005}]{d05}
Dud\'{i}k M, Schapire RE, Phillips SJ. (2005). 
Correcting sample selection bias in maximum entropy density estimation. 
Advances in Neural Information Processing Systems 17, 323–330.

\bibitem[\protect\citeauthoryear{Eguchi and Komori}{2022}]{e22}
Eguchi S, Komori O. (2022). 
Minimum Divergence Methods in Statistical Machine Learning: From an Information Geometric Viewpoint.
Springer Japan KK, part of Springer Nature, Tokyo.

\bibitem[\protect\citeauthoryear{Fithian~et~al.}{2015}]{f15}
Fithian W, Elith J, Hastie T, Keith DA. (2015). 
Bias correction in species distribution models: pooling survey and collection data for multiple species. 
Methods in Ecology Evolution 6, 424–438.

\bibitem[\protect\citeauthoryear{Fithian and Hastie}{2013}]{f13}
Fithian W, Hastie T. (2013) 
Finite-sample equivalence in statistical models for presence-only data. 
The Annals of Applied Statistics 7(4), 1917-1939.

\bibitem[\protect\citeauthoryear{Fujisawa and Eguchi}{2008}]{f08}
Fujisawa H, Eguchi S. (2008). 
Robust parameter estimation with a small bias against heavy contamination.
Journal of Multivariate Analysis 99, 2053-2081.

\bibitem[\protect\citeauthoryear{Goeman}{2010}]{g10}
Goeman JJ. (2010). 
L$_1$ penalized estimation in the Cox proportional hazards model. 
Biometrical Journal 52(1), 70-84.

\bibitem[\protect\citeauthoryear{Komori et~al.}{2015}]{k15}
Komori O, Eguchi S, Copas JB. (2015).
Generalized t-statistic for two-group classification.
Biometrics 71(2), 404-416.

\bibitem[\protect\citeauthoryear{Komori et~al.}{2020}]{k20}
Komori O, Eguchi S, Saigusa Y, Kusumoto B, Kubota Y. (2020).
Sampling bias correction in species distribution models by quasi-linear Poisson point process.
Ecological Informatics 55, 101015.

\bibitem[\protect\citeauthoryear{Kubota et~al.}{2015}]{k15b}
Kubota Y, Shiono T, Kusumoto B.  (2015).
Role of climate and geohistorical factors in driving plant richness patterns and endemicity on the east Asian continental islands.
Ecography 38(6), 639-648.

\bibitem[\protect\citeauthoryear{Lancaster and Imbens}{1996}]{l96}
Lancaster T, Imbens GW. (1996).
Case-control studies with contaminated controls.
Journal of Econometrics 70, 145-160.

\bibitem[\protect\citeauthoryear{Lele}{2009}]{l09}
Lele SR. (2009).
A new method for estimation of resource selection probability function. 
The Journal of Wildlife Management 73(1), 122-127.

\bibitem[\protect\citeauthoryear{Lele and Keim}{2006}]{l06b}
Lele SR, Keim JT. (2006).
Weighted distributions and estimation of resource selection probability functions. 
Ecology 87(12), 3021-3028.

\bibitem[\protect\citeauthoryear{Liu et~al.}{2018}]{l18}
Liu C, White M, Newell G. (2018).
Detecting outliers in species distribution data.
Journal of Biogeography 45, 164-176.

\bibitem[\protect\citeauthoryear{Meier et al.}{2008}]{m08}
Meier L, van de Geer S, Buhlmann P. (2008). 
The group lasso for logistic regression. 
Journal of the Royal Statistical Society, Series B 70(1), 53-71.

\bibitem[\protect\citeauthoryear{Mesibov}{2013}]{m13}
Mesibov R. (2013).
A specialist's audit of aggregated occurrence records. 
ZooKeys 293, 1-18.

\bibitem[\protect\citeauthoryear{M{\o}ller~et~al.}{1998}]{m98}
M{\o}ller J, Syversveen AR,  Waagepetersen RP. (1998). 
Log Gaussian Cox processes. 
Scandinavian Journal of Statistics 25(3), 451–482.

\bibitem[\protect\citeauthoryear{Ogata}{1978}]{o78}
Ogata, Y. (1978). 
The asymptotic behaviour of maximum likelihood estimators for stationary point processes. 
Annals of the Institute of Statistical Mathematics, 30(1), 243-261.

\bibitem[\protect\citeauthoryear{Phillips~et~al.}{2009}]{p09}
Phillips SJ, Dud\'{i}k M, Elith J, Graham CH, Lehmann A, Leathwick J, Ferrier S. (2009). 
Sample selection bias and presence-only distribution models: implications for background and pseudo-absence data. 
Ecological Applications 19(1), 181–197.

\bibitem[\protect\citeauthoryear{Rathbun and Cressie}{1994}]{r94}
Rathbun SL, Cressie N. (1994). 
Asymptotic properties of estimators for the parameters of spatial inhomogeneous Poisson point processes. 
Advances in Applied Probability, 26(1), 122-154.

\bibitem[\protect\citeauthoryear{Renner and Warton}{2013}]{r13}
Renner IW, Warton DI. (2013).
Equivalence of MAXENT and Poisson point process models for species distribution modeling in ecology.
Biometrics 69(1), 274–281. 

\bibitem[\protect\citeauthoryear{Schmeller~et~al.}{2018}]{s18}
Schmeller DS, Weatherdon LV, Loyau A, Bondeau A, Brotons L, Brummitt N, Geijzendorffer IR, Haase P, Kuemmerlen M, Martin CS,  Mihoub J-B, Rocchini D, Saarenmaa H, Stoll S and Regan EC. (2018).
A suite of essential biodiversity variables for detecting critical biodiversity change. 
Biological Reviews 93, 55–71.

\bibitem[\protect\citeauthoryear{Serra-Diaz~et~al.}{2017}]{s17}
Serra-Diaz JM, Enquist BJ, Maitner B, Merow C, Svenning J-C. (2007).
Big data of tree species distributions: how big and how good? 
Forest Ecosystems 4, 30.

\bibitem[\protect\citeauthoryear{Streit}{2010}]{s10}
Streit RL. (2010). 
Poisson Point Processes: Imaging, Tracking, and Sensing.
Springer, New York.

\bibitem[\protect\citeauthoryear{Thessen and Patterson}{2011}]{t11}
Thessen A, Patterson D. (2011).
Data issues in the life sciences. 
ZooKeys 150, 15–51.

\bibitem[\protect\citeauthoryear{Tibshirani}{1996}]{t96}
Tibshirani R. (1996). 
Regression shrinkage and selection via the lasso. 
Journal of the Royal Statistical Society, Series B 58(1), 267-288.

\bibitem[\protect\citeauthoryear{Vel\'{a}squez-Tibat\'{a}~et~al.}{2019}]{v19}
Vel\'{a}squez-Tibat\'{a} J, Olaya-Rodr\'{i}guez MH, L\'{o}pez-Lozano D, Guti\'{e}rrez C, Gonz\'{a}lez I, Londo\~{n}o-Murcia MC. (2019). 
BioModelos: A collaborative online system to map species distributions.
PLoS one 14, e0214522.

\bibitem[\protect\citeauthoryear{Veloz}{2009}]{v09}
Veloz SD. (2009). 
Spatially autocorrelated sampling falsely inflates measures of accuracy for presence-only niche models.
Journal of Biogeography 36, 2290–2299.

\bibitem[\protect\citeauthoryear{Wang and Stone}{2019}]{w19}
Wang Y, Stone L. (2019).
Understanding the connections between species distribution models for presence-background data.
Theoretical Ecology 12, 73-88.

\bibitem[\protect\citeauthoryear{Warton and Shepherd}{2010}]{w10}
Warton DI, Shepherd LC. (2010).
Poisson point process models solve the pseudo-absence problem for presence-only data in ecology.
The Annals of Applied Statistics 4, 1383–1402. 

\bibitem[\protect\citeauthoryear{Wiser}{2016}]{w16}
Wiser SK. (2016).
Achievements and challenges in the integration, reuse and synthesis of vegetation plot data. 
Journal of Vegetation Science 27, 868–879. 

\end{description}

\noindent
\appendix
\def\thesection{Appendix \Alph{section}}
\section{Derivation of estimating equation}
\label{appmide}

Let $\Xi(t)$ be a strictly convex function of $t$ defined on ($0,\infty$).
Then, the Bregman divergence between intensity functions $\lambda_1(s)$ and $\lambda_2(s)$ is induced by the generator function $\Xi$ as
\begin{eqnarray}
 D_\Xi(\lambda_1,\lambda_2) = \int_\mathscr{A}\left[ \Xi(\lambda_1(s))-\Xi(\lambda_2(s))-\xi(\lambda_2(s))\{\lambda_1(s)-\lambda_2(s)\} \right]ds,
\label{u-div}
\end{eqnarray}
where $\xi$ is the derivative of $\Xi$.
Note that $D_\Xi(\lambda_1,\lambda_2)\geq 0$ 
because the integrand of \eqref{u-div} is non-negative due to the convexity of $\Xi$.
The equality holds if and only if $\lambda_1=\lambda_2$.
If $\Xi(t)=t\log t-t$, then 
$$
D_\Xi(\lambda_1,\lambda_2)=\int [\lambda_1(s)\{\log(\lambda_1(s))-\log(\lambda_2(s))\}-\lambda_1(s)+\lambda_2(s)]ds,
$$
 which is the extended KL divergence defined on the space of intensity functions.

Let $\lambda_0(s)$ be the true intensity function.
We consider the minimum divergence estimation with the divergence $D_\Xi(\lambda_0,\lambda_{\bmt\theta})$ with respect to $\bm\theta$.
We note that
\begin{eqnarray}
D_\Xi(\lambda_0,\lambda_{\bmt\theta})=
-\int_{\mathcal{A}}\left[\Xi(\lambda_{\bmt\theta}(s))+\xi(\lambda_{\bmt\theta}(s))\{\lambda_0(s)-\lambda_{\bmt\theta}(s)\}\right]ds + C_1,
\end{eqnarray}
where $C_1$ is a constant that depends only on $\lambda_0$.
Hence we define the loss function as
\begin{eqnarray}
l_\Xi(\bm\theta) = 
-  \sum_{i=1}^m\xi(\lambda_{\bmt\theta}(s_i)) + \int_\mathscr{A}\left\{ \lambda_{\bmt\theta}(s)\xi(\lambda_{\bmt\theta}(s)) - \Xi(\lambda_{\bmt\theta}(s)) \right\}ds,
\end{eqnarray}
which has a numerical approximation as
\begin{eqnarray}
  l_\Xi(\bm\theta) = 
  - \sum_{i=1}^r\left[ d_i\xi(\lambda_{\bmt\theta}(s_i)) - w_i\lambda_{\bmt\theta}(s_i)\xi(\lambda_{\bmt\theta}(s_i)) + w_i\Xi(\lambda_{\bmt\theta}(s_i)) \right] \label{lossb}
\end{eqnarray}
using the presence indicators $d_i$'s and the quadrature weights $w_i$'s as in Section 2.
We note that the expectation under the true intensity function $\lambda_0(s)$ is equal to  $D_\Xi(\lambda_0,\lambda_{\bmt\theta})$ except for the constant term $C_1$. 

Let $F$ be a cumulative distribution function defined on $(0,\infty)$.
Then, apply the general discussion for the divergence $D_\Xi$ to a specific generator function $\Xi$ that is defied by
\begin{eqnarray}
\Xi(t)=\int_0^t\int_0^s\frac{F(\tau u)}{u}duds
\end{eqnarray}
Accordingly, $\Xi$ is a convex function because  $(d^2/dt^2) \Xi(t)=F(\tau t)/t$ is nonnegative.
For the model intensity $\lambda_{\bmt\theta}=\lambda_{\bmt\beta}b_{\bmt\alpha}$, the estimating functions in (\ref{eeb}) and (\ref{eeb2}) are given as the derivatives of the loss function $l_\Xi(\bm\theta)$ with respect to $\bm\beta$ and $\bm\alpha$, that is
\begin{align}
\frac{\partial}{\partial\bm\beta}l_\Xi(\bm\theta)&=\sum_{i=1}^rF(\tau\lambda_{\bmt\theta}(s_i))\{ d_i-w_i\lambda_{\bmt\theta}(s_i) \}\bm x(s_i), \\
\frac{\partial}{\partial\bm\alpha}l_\Xi(\bm\theta)&=
\sum_{i=1}^r\frac{F(\tau\lambda_{\bmt\theta}(s_i))}{1+\exp\left(\bm\alpha^\top \bm z(s_i)\right)}\{ d_i-w_i\lambda_{\bmt\theta}(s_i) \}\bm z(s_i).
\end{align}

\section{Consistency and asymptotic normality of the MIDE}
\label{appmide2}

We follow the asymptotic results described in \citet{o78,r94,a99}.
The estimating functions based on loss function (\ref{lossb}) is unbiased because
\begin{eqnarray}
  E_{\lambda_{\bmt\theta}}\left[\frac{\partial}{\partial\bm\beta}l_\Xi(\bm\theta)\right] = \hspace{-1em}&& - \int_\mathscr{A}\left\{\xi^\prime(\lambda_{\bmt\theta}(s))b_{\bmt\alpha}(s)\frac{\partial}{\partial\bm\beta}\lambda_{\bmt\beta}(s)\right\}\lambda_{\bmt\theta}(s)ds \nonumber \\  && + \int_\mathscr{A}\lambda_{\bmt\theta}(s)\xi^\prime(\lambda_{\bmt\theta}(s))b_{\bmt\alpha}(s)\frac{\partial}{\partial\bm\beta}\lambda_{\bmt\beta}(s)ds = \bm 0_{1+p}, 
\end{eqnarray}
where $\xi^\prime$ is the derivative of $\xi$, and the expectation is taken with respect to PPP with the model intensity function. 
Define $\bm A_t=\{ t\bm a, \bm a\in \bm A \}$ where $\bm A\subset \mathbb R^2$ be a compact set with positive Lebesgue measure and assume that $\bm A_t\uparrow \mathbb R^2$ as $t\to\infty$.
Consider the MIDE based on a PPP on $\bm A_t$.
Let $\bm\beta_0$ denote the solution of $E_{\lambda_0}\left[ (\partial/\partial\bm\beta)l_\Xi(\bm\theta) \right]=\bm 0_{1+p}$, where the expectation is taken with respect to PPP with the true intensity function.
Under some regularity conditions for the proof of Theorems 1 and 2 in \citet{a99}, $\hat{\beta}_\tau\xrightarrow{p}\beta_0$ and
\begin{eqnarray}
\sqrt{\Lambda_\theta}\left( \hat{\bm\beta}_\tau-\bm\beta_0 \right)\xrightarrow{d} N\left( \bm 0, \bm\Sigma_\tau(\bm\beta_0) \right)
\end{eqnarray}
as $t\to\infty$,  where $\Lambda_\theta=\int_{\bmt A_t}\lambda_\theta(s)ds$ and $\bm\Sigma_\tau(\bm\beta_0)=\bm J_\tau(\bm\beta_0)^{-1}\bm I_\tau(\bm\beta_0)\bm J_\tau(\bm\beta_0)^{-1}$,
\begin{align}
\bm J_\tau(\bm\beta) &= \frac1{\Lambda_\theta}E_{\lambda_0}\left[ F\left( \tau\lambda_\theta(s) \right)\bm x(s)\bm x^\top(s)  \right], \\
\bm I_\tau(\bm\beta) &= \frac1{\Lambda_\theta}E_{\lambda_0}\left[ F\left( \tau\lambda_\theta(s) \right)^2\bm x(s)\bm x^\top(s)  \right].
\end{align}

\section{Gradient ascent algorithm}
\label{appga}

The gradient ascent method is employed to minimize the penalized loss.
We follow the approach described in \citet{m08,k15}.
Consider the gradient $\bm g(\bm\beta|\bm\alpha)=(g_0(\bm\beta|\bm\alpha),g_1(\bm\beta|\bm\alpha),\dots,g_p(\bm\beta|\bm\alpha))^\top$ defined by
\begin{equation}
g_k(\bm\beta|\bm\alpha)=
    \begin{cases}
        \displaystyle \frac{\partial}{\partial \beta_k}l_\Xi(\bm\theta) - \phi_k\sign(\beta_k)   &  {\rm if}\ \beta_k\neq 0  \\
        \displaystyle \frac{\partial}{\partial \beta_k}l_\Xi(\bm\theta) - \phi_k\sign\left( \frac{\partial}{\partial \beta_k}l_\Xi(\bm\theta) \right)   &  \displaystyle {\rm if}\ \beta_k=0\ {\rm and}\ \left| \frac{\partial}{\partial \beta_k}l_\Xi(\bm\theta) \right|>\phi_k \\
        0  &  {\rm otherwise}
    \end{cases}
\end{equation}
for $k=0,1,\dots,p$, where $\phi_0=0$ and $\phi_k=\phi$ for $k\neq 0$ and `sign' indicates the sign function.
To find an optimal gradient within a subdomain on which the gradient is continuous, consider the range defined by
\begin{eqnarray}
\rho_{\rm edge}(\bm\beta) = \min_{k=1,\dots,p}\left[ -\frac{\beta_k}{g_k(\bm\beta|\bm\alpha)}\middle| \sign(\beta_k)=-\sign(g_k(\beta_k))\neq 0 \right].
\end{eqnarray}
The unpenalized parameter $\bm\alpha$ is estimated by combining the Newton-Raphson method.
Given values of tuning parameters $\tau$ and $\phi$, an iterative estimation procedure is described as follows:

\noindent\hrulefill
\begin{enumerate}
\item Initialize $\bm\beta^{(1)}$ and $\bm\alpha^{(1)}$.
\item For steps $t=2,3,\dots$, iteratively calculate  
\begin{align}
\bm\beta^{(t)}&=\bm\beta^{(t-1)}+\rho_{\rm opt}\bm g\left(\bm\beta^{(t-1)} \middle| \bm\alpha^{(t-1)}\right), \\
\bm\alpha^{(t)}&=\bm\alpha^{(t-1)}-{\bm H}_{\bmt\alpha}^{-1}\left(\bm\alpha^{(t-1)} \middle| \bm\beta^{(t)}\right){\bm g}_{\bmt\alpha}\left(\bm\alpha^{(t-1)} \middle| \bm\beta^{(t)}\right),
\end{align}
where $\bm g_{\bmt\alpha}$ and $\bm H_{\bmt\alpha}$ are the gradient and Hessian of $l_\Xi^\phi$ with respect to $\bm\alpha$, respectively, and
\begin{eqnarray}
\rho_{\rm opt} = \argmin_{0\leq\rho\leq\rho_{\rm edge}(\bmt\beta^{(t-1)})}l_\Xi^\phi\left\{ \bm\beta^{(t-1)}+\rho\bm g\left(\bm\beta^{(t-1)}\middle| \bm\alpha^{(t-1)}\right) \right\}
\end{eqnarray}
until convergence with respect to the penalized loss.
\end{enumerate}
\hrulefill

\noindent
This procedure is repeated when using different values of $\phi$, $\phi_{(n_\phi)}>\cdots>\phi_{(1)}=0$, where $\phi_{(n_\phi)}=\max_{1\leq k\leq p}|\partial l_\Xi(\bm\beta^{(1)})/\partial \beta_k|$ and $n_\phi$ is appropriately set according to $p$.
The starting value $\bm\beta^{(1)}$ when using $\phi_{(u)}$ is set at the resultant estimate when using $\phi_{(u-1)}$ for $u=2,\dots,n_\phi$.



\end{document}